\documentclass[apj]{emulateapj}

\newcommand{\hmpc}{\ifmmode{h^{-1}\,\hbox{Mpc}}\else{$h^{-1}$\thinspace Mpc}\fi}
\newcommand{\etal}{et~al.}
\newcommand{\kms}{\ifmmode{\,\hbox{km}\,s^{-1}}\else {\rm\,km\,s$^{-1}$}\fi}
\newcommand{\msun}{{\rm\,M_\odot}}

\begin{document}
\title{Clustering of supernova {\rm Ia} host galaxies}
\shortauthors{Carlberg \etal}
\author{
R.~G.~Carlberg\altaffilmark{1},
M.~Sullivan\altaffilmark{1,2},
D.~Le~Borgne\altaffilmark{3}, 
A.~Conley\altaffilmark{1},
D.~A.~Howell\altaffilmark{1},
K.~Perrett\altaffilmark{1},
P.~Astier\altaffilmark{4}, D.~Balam\altaffilmark{5}, 
C.~Balland\altaffilmark{4},S.~Basa\altaffilmark{6}, 
D.~Hardin\altaffilmark{4},D.~Fouchez\altaffilmark{4}, J.~Guy\altaffilmark{4}, I.~Hook\altaffilmark{2}, R.~Pain\altaffilmark{4}, C.~J.~Pritchet\altaffilmark{5}, N.~Regnault\altaffilmark{4}, J.~Rich\altaffilmark{3}, S.~Perlmutter\altaffilmark{7}
 }

\altaffiltext{1}{Department of Astronomy and Astrophysics, University
  of Toronto, Toronto, ON M5S 3H4, Canada}
\altaffiltext{2}{University of Oxford Astrophysics, Denys Wilkinson
  Building, Keble Road, Oxford OX1 3RH, UK}
\altaffiltext{3}{DSM/IRFU, CEA/Saclay, 91191 Gif-sur-Yvette Cedex, France}
\altaffiltext{4}{LPNHE, CNRS-IN2P3 and University of Paris
  VI \& VII, 75005 Paris, France} 
\altaffiltext{5}{Department of Physics and Astronomy, University of
  Victoria, PO Box 3055, Victoria, BC V8W 3P6, Canada}
\altaffiltext{6}{LAM CNRS,
  BP8, Traverse du Siphon, 13376 Marseille Cedex 12, France}
\altaffiltext{7}{Lawrence Berkeley National Laboratory, 1 Cyclotron
  Rd., Berkeley, CA 94720, USA}

\email{carlberg@astro.utoronto.ca }

\begin{abstract}

For the first time the cross-correlation between type Ia supernova host galaxies and surrounding field galaxies is measured using the Supernova Legacy Survey sample. Over the z=0.2 to 0.9 redshift range we find that supernova hosts are correlated an average of 60\% more strongly than similarly selected field galaxies over the $3-100\arcsec$ range and about a factor of 3 more strongly below $10\arcsec$. The correlation errors are empirically established with a jackknife analysis of the four SNLS fields. The hosts are more correlated than the field at a significance of 99\% in the fitted amplitude and slope, with the point-by-point difference of the two correlation functions having a reduced $\chi^2$ for 8 degrees of freedom of 4.3, which has a probability of random occurrence of less than $3\times 10^{-5}$. The correlation angle is $1.5\pm 0.5\arcsec$, which deprojects to a fixed co-moving correlation length of approximately $6.5\pm 2\hmpc$. Weighting the field galaxies with the mass and star formation rate supernova frequencies of the simple A+B model produces good agreement with the observed clustering. We conclude that these supernova clustering differences are primarily the expected outcome of the dependence of supernova rates on galaxy masses and stellar populations with their clustering environment. 
\end{abstract}


\section{INTRODUCTION}
\nobreak
The relationship between clustering environment, supernova production rates and subsequent feedback to the environment has long been of interest. The early investigations of the relation between Ia supernova rates and clustering were based on local samples largely targeted towards clusters. A study of 40 some nearby supernovae primarily in cluster fields concluded that field and cluster rates were not distinguishable \citep{Zwicky:42, Barbon:68}. Later, when the host galaxy luminosity dependence of supernova production frequencies was recognized, the cluster rate was found to be lower than the field \citep{Barbon:78}.  Most supernovae in those samples are likely Type Ia although the data and supernova typing methods of the time were not able to make a clean separation between core-collapse supernovae and white dwarf explosions. The direct dependence of the supernova rate on luminosity, or total stellar mass and the star formation rate was remarked on in a number of different contexts \citep{vdB:59, vdB:60, Tammann:70, CTW:77, CO:81}. One recent search targeted more than one hundred Abell clusters found six new supernova Ia and concluded that clusters had a supernova rate similar to the field \citep{Sharon:07}. A recent analysis of a sample of 136 local supernova reached the conclusion that cluster early type galaxies have a supernova Ia rate about three times the rate in field early types at 98\% significance \citep{Mannucci:07}. The application of supernovae as cosmological distance indicators is subject to the concern that the intrinsic luminosities are evolving. These clustering studies raise the possibility that supernovae, and especially those in elliptical galaxies, may be subject to environmental influences. This paper first undertakes a new evaluation of the influence of clustering and then tries to determine whether any differences can be reasonably explained as the result of host galaxy population differences with clustering environment, i.e. the well known tendency for relatively more early type galaxies to be present in clusters.

The Supernova Legacy Survey (SNLS) sample probes out to redshift one, a range which contains substantial evolution of the host galaxy and supernova population which provides guidance to processes at higher redshifts when galaxies were undergoing their primary star formation.  Supernova Ia are of particular interest because they dominate the production of iron peak elements and inject some $10^{51}$ erg of kinetic energy starting approximately $10^8$ years after their progenitor stars were formed. A field supernova survey like SNLS offers a fair sample over all clustering environments but has the statistical difficulty that only a few percent of all galaxies are in clusters, leading to relatively small numbers even in relatively large samples like the SNLS \citep{Astier:06, Melissa:07}. However, the SNLS has the density and depth of both supernova hosts and field galaxies that we can undertake a cross-correlation analysis that takes advantage of the approximately 280 identified host galaxies in our current 3-year sample. This type of analysis does not separate the environment into field, group or cluster, but usefully expresses the statistical average over all environments and complements analyses which specifically look at clusters.

A simple, testable, theory relating supernova hosts to field galaxies is available. Quite generally the rate of production of Ia supernovae is the convolution of the star formation rate with $D(t-t_f)$,the fractional distribution of delay time between the formation of the stars at $t_f$ and that fraction that explode as Ia at $t$, or, $R(t)= \int_0^t D(t-t_f) \dot{M}(t_f) dt_f$. The delay time distribution can, in principle, be calculated given the evolution and (largely unknown) frequencies of close binaries into supernova Ia \citep{Greggio:05}. Scannapieco \& Bildsten(2005) introduced the simple empirical approximation that breaks the delay time distribution into a two piece discontinuous model having a prompt component proportional to a star formation indicator and a delayed component proportional to the built up stellar mass of the galaxy. That is, $D(t-t_f)$ is $A$ for $[0,t-\Delta t]$, and $B/\Delta t$ for $[t-\Delta t, t]$. Doing the integrals gives the resulting supernova Ia rate,  as $R(Ia) = A M + B \langle\dot M\rangle_{\Delta t}$, where the stellar mass, $M$, and recent star formation rate, $\langle\dot{M}\rangle_{\Delta t}$ (abbreviated to $\dot{M}$) can both be inferred from our data. The model can be fit to the host and field galaxies at a single redshift, but then predicts a redshift dependence of the rates \citep{Sullivan:06, Mannucci:07, Howell:07} and the clustering dependence. In this paper we will test whether this weighting of an appropriate field sample can reproduce the clustering of supernova Ia hosts. 

\section{DATA and METHODS}
\label{sec:sample}
\nobreak 
The identifications of the supernova host galaxies and the mass and star formation rates of all galaxies are derived with the methods of Sullivan \etal\ (2006). In brief, we have deep [$u_M,g_M,r_M,i_M,z_M$] images of the four SNLS fields in the Megaprime filter system (similar to the SDSS filters). From the photometry we derived indicative star formation rates, stellar masses and photometric redshifts for all of the galaxies using the PEGASE approach \citep{LeBorgne:02, LeBorgne:04}. The SNLS sample used here includes detections up to early 2007 with 281 spectroscopically confirmed supernova Ia host galaxies. To produce a sample suitable for clustering studies, we first ensure that field and hosts cover the same observational space in brightness, redshift, and mass and on the sky. The first cuts are simply to remove supernovae that have no clear host or are in regions of the array that are masked out, usually because of the presence of a bright star. Second, we restrict the mass of the host galaxy to be in the range of $10^8 \msun$ to $10^{11.55}\msun$ which eliminates about twenty low mass hosts, but none at high mass. The main benefit is to remove about half of the field sample and hence reduce the noise in the measurement. We restrict the galaxy brightness range to $20\le i_M\le 25$ mag, Figure~1, and the redshifts to be in the range of 0.2 to 0.9, Figure~2. The sample starts to become incomplete around redshift 0.6, but this is not a problem for a correlation analysis. The cuts leave a total of 163 supernova host galaxies and 147,246 field galaxies in the four fields. The 46 hosts in the range $18\le i_M \le 20$ mag are at low redshift and complicate the problem of producing a comparable reference population from the field galaxies, which are mainly at higher redshifts.

\begin{figure}
\plotone{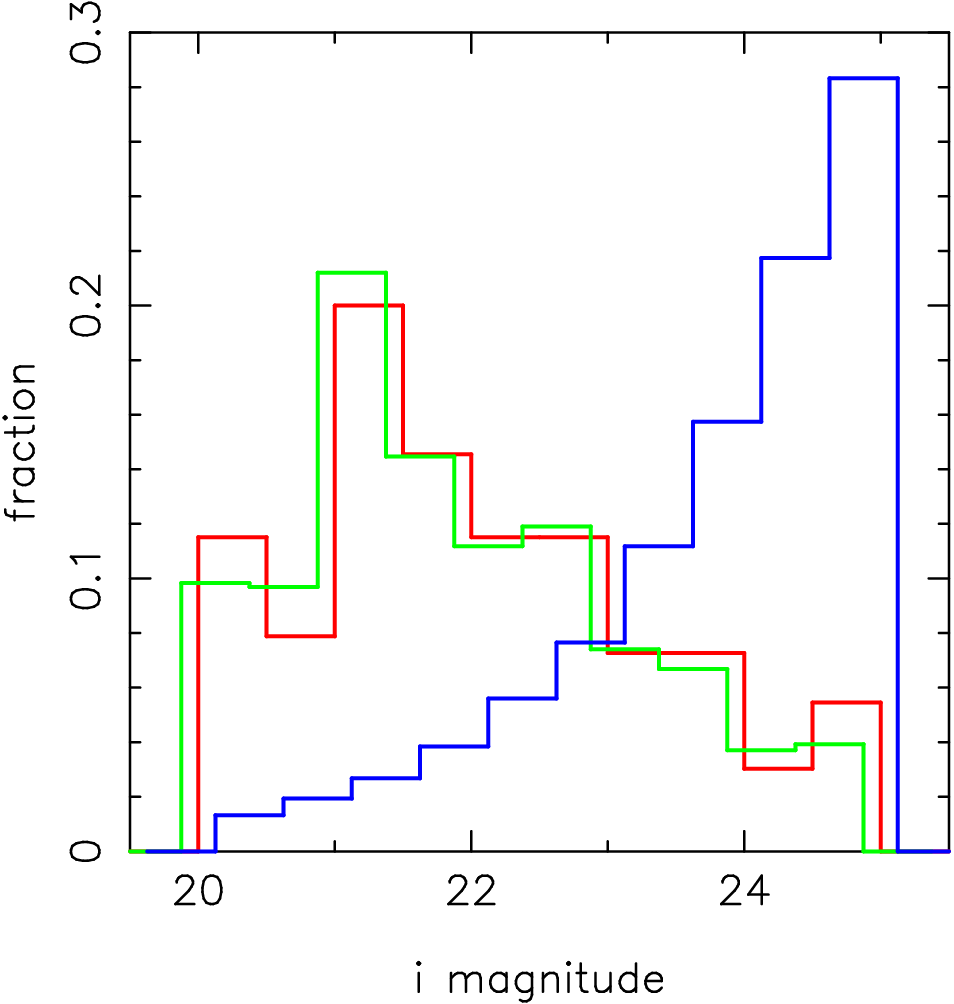}
\caption{The $i_M$ magnitude distribution of host galaxies (solid/red), imitation hosts (dashed/green, offset left), and field galaxies (dotted/blue, offset right). Each histogram area in the 0.5 mag bins is normalized to unity.
\label{fig_m}}
\end{figure}

\begin{figure}
\plotone{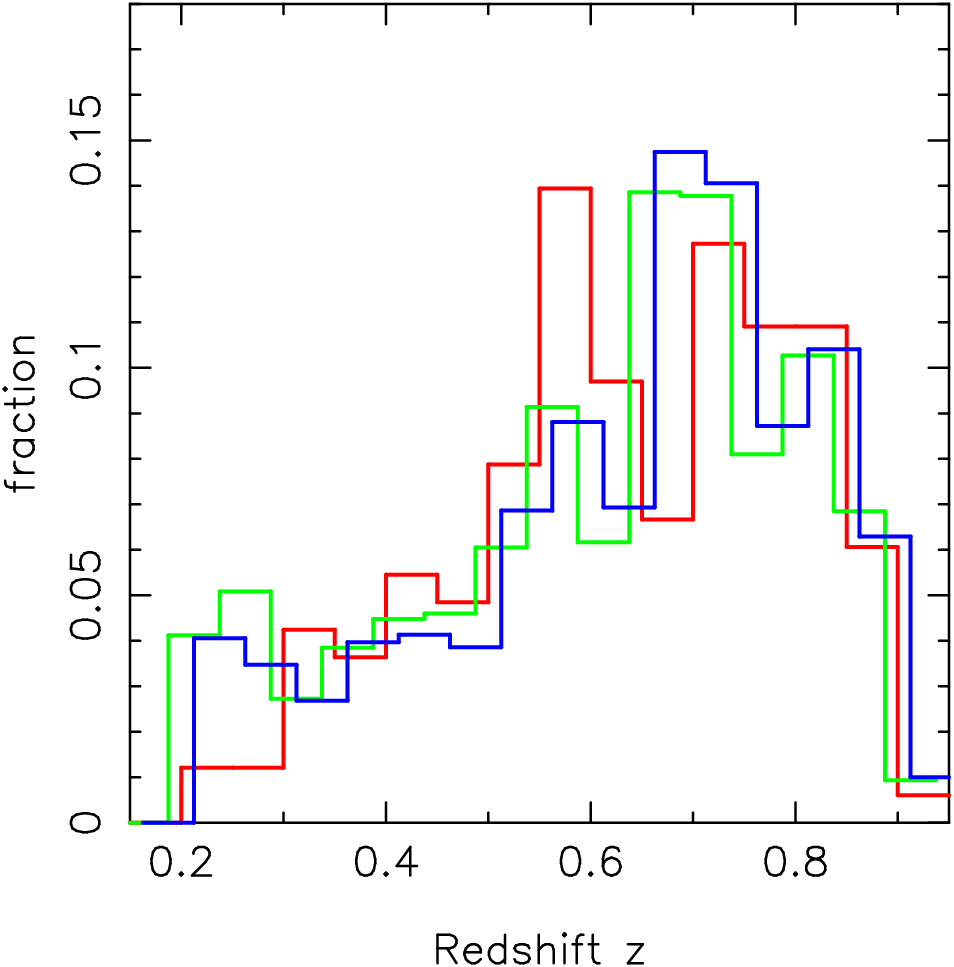}
\caption{The redshift distribution of host galaxies (solid/red), imitation hosts (dashed/green, offset left), and field galaxies (dotted/blue, offset right). Each histogram areas in the $\Delta z=0.05$ bins is normalized to unity. 
\label{fig_z}}
\end{figure}

We will measure the clustering of galaxy hosts relative to the field with the angular two-point cross-correlation function. The two-point function gives the sky density of field galaxies around a supernova Ia host galaxy at an angular separation $\theta$ as $n_{hf}(\theta) = n_f[1+w_{hf}(\theta)]$. For galaxy magnitudes in the range of an $i_M \simeq 25$ mag clustering of field galaxies leads us to expect that the $w_{hf}(\theta)$ will be generally less than one beyond a few arc-seconds, that is, over the entire range we probe. To estimate these weak correlations we use the cross-correlation form of the Landy-Szalay (1993) estimator modified to have the required symmetry in the two populations (Blake \etal 2006). 

The random sample is generated by uniformly placing $10^5$ points on the sky in each field and then removing the points that fall in the masked out regions. This leads to about 90,000 points per field, about three times the number of field galaxies, which is sufficient to reduce shot noise errors well below field to field variance. We take advantage of the spectroscopic redshifts of the supernovae hosts and the photometric redshifts of the field to boost the signal to noise of the clustering measurement by requiring that the field galaxy be within $\pm 0.1$ in redshift of the host spectroscopic redshift. The selected redshift window approximately matches the errors in the photometric redshifts \citep{Sullivan:06}. Other values were tried, but the selected 0.1 is close to optimal, giving about a factor of two improvement over no redshift selection at all. 

To compare the clustering of supernovae hosts to field galaxies, we construct an imitation host sample distributed in redshift and brightness identically to the real hosts. We randomly draw from the field galaxies a sample of 10000 galaxies in each field with the same $i_M$ distribution as the hosts, Figure~\ref{fig_m}. The figure shows that the supernova hosts have a very broad flat brightness distribution compared to the rising counts of the field brightness distribution. Once the magnitude selection is done, the redshift distribution of the imitation hosts is sufficiently similar to the host galaxy redshift distribution as shown in Figure~\ref{fig_z} that we do not futher select to improve the imitation hosts. An important aspect of this sample is that the precision of the clustering measurement derived from it is completely dominated by the structures in the 4 fields. To quantify this, we reduced the sample to 3000 imitation hosts per field, finding that the errors have increased by 13\%, far short of the $\sqrt{3}$ expected from sample size dominated Poisson statistics. We conclude that for $10^4$ imitation hosts the Poisson errors are about a 10\% component of the errors, with the field to field component being in common to both samples. 

We will compare the host-field cross-correlation with the imitation host-field cross-correlation. The imitation hosts are compared with unit weights and with weights assigned from the A+B formulation which reproduces the expected supernova frequencies.  The A and B values from \citep{Sullivan:06} are adopted. In principle these are not quite appropriate, since the hosts here are not weighted for detection efficiency. We did determine A and B for the sample used here, finding no practical difference in the outcome.

\section{RESULTS}

\begin{figure}
\plotone{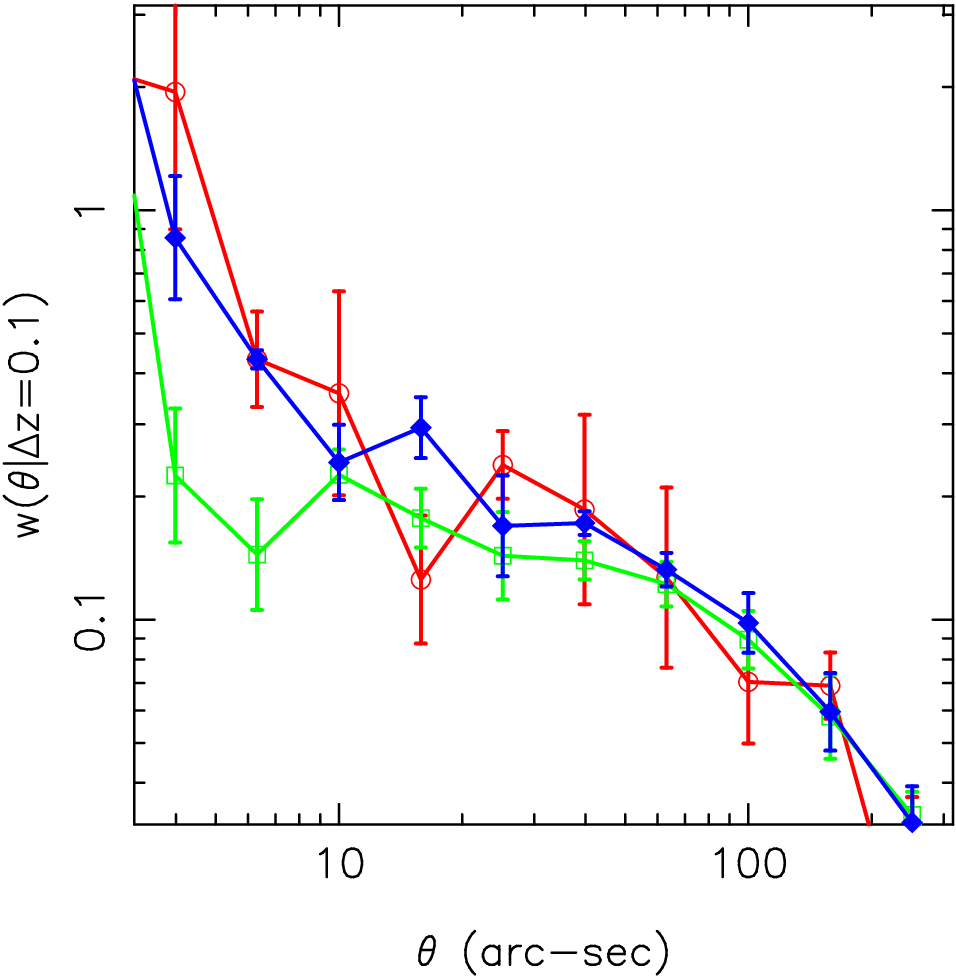}
\caption{The log-log plot of the angular cross-correlation between the field galaxies and the supernova hosts (dashed/red, open circle), the imitation hosts population (dashed-dotted/green, open square) and the A+B weighted imitation hosts (dotted/blue, diamonds). 
\label{fig_w}}
\end{figure}

Figure~\ref{fig_w} shows the angular cross-correlation of the field and hosts in the redshift range 0.2 to 0.9, with all galaxies required to have $20\le i_M \le 25$ mag and $8\le \log{M_\ast} \le 11.55$. On the average, the host-field cross-correlation is slightly stronger than the comparably selected field galaxies, with the difference being largest at separations closer than about $10\arcsec$. To quantify the cross-correlation we fit the power law model $w(\theta) = A_w \theta^{-\delta}$, equivalent to $(\theta_0/\theta)^{-\delta}$, where $\theta_0$ is the correlation angle. The fitting is done as a linear fit in $\log{w}-\log{\theta}$. The error at any point is estimated with the Jackknife procedure, in which the difference of the results recalculated dropping one field at a time and all four fields are summed in quadrature and divided by $n-1$ to estimate the variance. The jackknife approach gives realistic, unbiased errors.  We exclude measurements inside $3\arcsec$ on the basis that the galaxies begin to physically overlap and it removes the self-correlation component. We fit over the $3-100\arcsec$ range where the correlations are always positive.
On the average the correlation of the real hosts is 60\% stronger over the entire radial range, a full factor of three more in the $3-10\arcsec$ range about about 12\% stronger over the $10-100\arcsec$ range.

\begin{figure}
\plotone{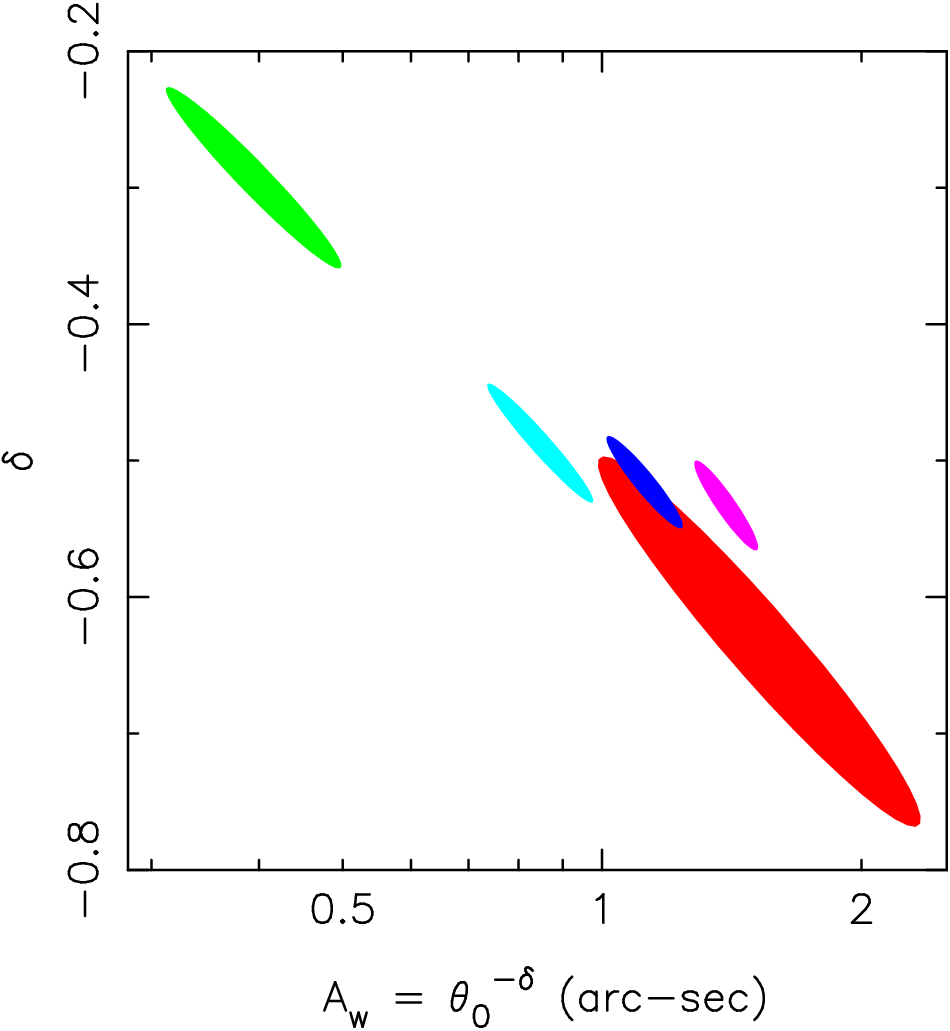}
\caption{The fit of the measured correlations to $A_w=\theta_0^{-\delta}$ with $\theta$ measured in arc-seconds. Power law fits over the range $3 -100\arcsec$ are shown for the supernova hosts (red, open circle), imitation hosts (green, open square), and A+B weighted imitation hosts (blue, diamond), A=0 (cyan, filled circle) and B=0 magenta, hexagon). The error ellipse is computed from the co-variance matrix of the fit.}
\label{fig_ga}
\end{figure}

\begin{table}
\begin{center}
\caption{Power Law Cross-Correlation Fits\label{tbl}}
\begin{tabular}{lrrrr}
\tableline\tableline
Population & $\log{A_w}$ & $\delta$  & $\chi^2_8$ fit & $\chi^2_8$ data \\
\tableline
hosts & 0.182 & -0.633 & 1.00 & 0.0 \\
imitation hosts & -0.404 & -0.293 &2.64 & 4.26 \\
A+B wts         & 0.049 & -0.516 & 1.11 & 1.43 \\
A=0 wts & -0.072 & -0.487 & 1.32 & 2.25 \\
B=0 wts & 0.144 & -0.533 & 1.47  & 1.50 \\
\tableline
\end{tabular}
\end{center}
\end{table}

Table 1 gives the best fit $[\log{A_w},\delta]$ values for the power-law fit to $3-100\arcsec$ range of the angular two-point correlation functions in the first two columns.  The fitted parameters are shown in Figure~4, along with their error ellipses, which show the usual strong correlation between the amplitude and slope. 
The third column of Table~1 gives the reduced $\chi^2$ of the cross-correlations when modelled with the host field galaxy fit in the first row. The true host fit is rejected as inconsistent with the imitation hosts at 99.3\% confidence. A better measure of the significance of the clustering difference between the hosts and the field is the $\chi^2$ difference of the two correlations functions. As discussed above we calculate $\chi^2$ using the jackknife computed variance of the host-field cross-correlations which includes both the Poisson errors and field-to-field variance which is in common, as discussed above. These values are presented in the last column of Table~1. We see that the hosts and imitation hosts have a $\chi^2$ per 8 degrees of freedom of 4.3, which has a probability of occurrence less than $3\times 10^{-5}$. The table also presents the results for \citep{Sullivan:06}the A+B weighting of the host galaxies in the correlation function. Weighting with star formation or mass alone, $A=0$, partially explains the effect, but the reduced $\chi^2$ of the difference between the correlation functions indicates that the agreement is not statistically adequate. Mass weighting alone does provide an acceptable fit to the clustering pattern. We conclude that the A+B weights work to account for the rate of supernovae in strongly clustered, relatively high mass, galaxies. 

To estimate the physical correlation length we use the Limber equation. The photometric redshift distribution is adequately fit with $n(z)$ proportional to $z \exp{(-{1\over 2}[(z-\bar{z})/\sigma_z]^2)}$ with $\bar{z}=0.6$ and $\sigma_z=0.18$. This distribution is convolved with a Gaussian with a redshift dispersion of 0.1. We find that the host-field correlation angle of approximately $1.5\pm 0.5\arcsec$ with $\delta=-0.63\pm 0.13$ corresponds to an (assumed constant) co-moving correlation length of $6.5\pm 2\hmpc$ in a flat $\Omega_M=0.24$ cosmological model. It must be borne in mind that this cross-correlation applies to a specific field sample, spanning the $10^8-10^{11.55}\msun$ mass range. The correlation length is marginally stronger than the canonical $5\hmpc$ of typical luminosity galaxies, and bolsters the view that the relatively more massive galaxies dominate the supernova clustering.

\section{DISCUSSION}

The main outcome of this paper is the first measurement of the two-point correlation between supernova hosts and the surrounding field galaxies. Over this redshift range we find that the hosts are more correlated than a similar $i_M$ magnitude and redshift selected field sample with the probability that this is a chance event of $3\times 10^{-5}$. Weighting the field with $AM_\ast + B \dot{M}$ reproduces the host clustering remarkably well, primarily as a result of accounting for the higher mass-to-light of older stellar populations. Mass weighting alone can explain the clustering properties for this sample as shown in Table~1, although it is important to note that the supernova rates at these redshifts are completely dominated by the star formation rates, not the mass, of the galaxies. The star formation rate is correlated with the galaxy mass which partially explains why either mass or star formation weighting alone works, but the very strong correlations of the most massive galaxies, which have effectively no star formation but a significant supernova Ia production, is also part of the effect. We conclude that supernova Ia host galaxy clustering is primarily the outcome of the dependence of galaxy stellar populations on clustering environment. 

\acknowledgements

This paper is based on observations obtained with MegaPrime/MegaCam, a joint project of CFHT and CEA/DAPNIA, at the Canada-France-Hawaii Telescope (CFHT) which is operated by the National Research Council (NRC) of Canada, the Institut National des Sciences de l'Univers of the Centre National de la Recherche Scientifique (CNRS) of France, and the University of Hawaii. This work is based in part on data products produced at the Canadian Astronomy Data Centre as part of the CFHT Legacy Survey, a collaborative project of NRC and CNRS.  Canadian collaboration members acknowledge support from NSERC and CIFAR; French collaboration members from CNRS/IN2P3, CNRS/INSU and CEA. MS acknowledges support from the Royal Society.
This research has made use of the NASA/IPAC Extragalactic Database (NED)which is operated by the Jet Propulsion Laboratory, California Institute of Technology, under contract with the National Aeronautics and Space Administration. 

{\it Facilities:} \facility{CFHT}, \facility{Gemini}, \facility{VLT}, \facility{Keck}.

\end{document}